\documentclass{sf2a-conf2014}
\usepackage{graphicx}
\usepackage{sidecap}
\usepackage{hyperref}
\usepackage{amssymb,amsmath}
\usepackage[]{natbib}  
\usepackage{epstopdf}

\def\BibTeX{{\rm B\kern-.05em{\sc i\kern-.025em b}\kern-.08em
    T\kern-.1667em\lower.7ex\hbox{E}\kern-.125emX}}
\bibpunct{(}{)}{;}{a}{}{,}  

\begin{document}

\TitreGlobal{SF2A 2014}


\title{Dynamics of exoplanetary systems, links to their habitability}

\runningtitle{Dynamics of exoplanetary systems, links to their habitability}

\author{E. Bolmont}\address{Univ. Bordeaux, LAB, UMR 5804, F-33270, Floirac, France \\
   			  CNRS, LAB, UMR 5804, F-33270, Floirac, France}
\author{S.\,N. Raymond$^{1}$}
\author{F. Selsis$^{1}$}

\setcounter{page}{237}


\maketitle


\begin{abstract}
Our knowledge of planets' orbital dynamics, which was based on Solar System studies, has been challenged by the diversity of exoplanetary systems.

Around cool and ultra cool dwarfs, the influence of tides on the orbital and spin evolution of planets can strongly affect their climate and their capacity to host surface liquid water.

We illustrate the role of tides and dynamics with the extreme case of planets orbiting around brown dwarfs. In multiple planet systems, the eccentricity is excited by planet-planet interactions. Planets are therefore heated up from the inside by the tidally-induced friction. This process can heat a habitable zone planet to such a level that surface liquid water cannot exist.

We also talk about the newly discovered potentially habitable Earth-sized planet Kepler-186f. Given the poorly estimated age of the system, the planet could still be evolving towards synchronization and have a high obliquity or be pseudo-synchronized with a zero obliquity. These two configurations would have a different effect on the climate of this planet.

\end{abstract}



\section{Introduction}
Since 1995 and the discovery of the first exoplanet orbiting a Sun-like star \citep{MayorQueloz1995}, the number of detected exoplanets has been steadily increasing. 
With now almost 1500 confirmed exoplanets (\url{http://exoplanets.org/}) and around 20 of them good candidates to host surface liquid water (\url{http://phl.upr.edu/projects/habitable-exoplanets-catalog}), we enter in an fascinating age. 

With the improvements made in exoplanet detection, we are able to detect planets less and less massive (or smaller and smaller) approaching the mass (or size) range of the Earth. We are now also able to probe the habitable zone of stars. We define here the habitable zone (HZ) as the region around a star where a planet with the right atmosphere can potentially sustain surface liquid water \citep{Kasting.etal:1993,Selsis2007}. 

Most of the detected HZ planets are either too massive (radial velocity planets) or too large (transit planets) to be categorized unequivocally as terrestrial planets. For example, Kepler-22b, a 2.4~$R_\oplus$ planet could be either a mini-Neptune or a super-Earth \citep{Borucki.etal:2012}. 
However \citet{Quintana2014} reports the discovery of the first Earth-sized planet in the HZ of a low mass star: this planet has a radius only 10\% bigger than Earth's. This planet is the closest we know to Earth, it is very likely rocky and can potentially host surface liquid water.

\smallskip

Unfortunately, being in the HZ of a star does not mean that the planet hosts surface liquid water. The presence of water depends on many different physical parameters and quantities. Not only does it depend on the characteristics of the atmosphere (pressure, temperature) and of its composition but also on the dynamics of the orbit of the planet. Indeed, the eccentricity of the orbit as well as the rotation period of the planet and its obliquity (the angle between the rotation axis and angular momentum vector) have an effect on the climate of a planet \citep[the so-called Milankovitch cycles that govern the glaciation periods on Earth;][]{berger1988}. As all these quantities are influenced by tidal forces, one should take into account the tidal orbital evolution of the system in order to assess the potential of a planet to host water.

In Section \ref{tides}, we describe how tides do influence the parameters affecting the climate of a planet. In Section \ref{BDs}, we study the evolution of planets around brown dwarfs which constitutes an extreme case of how dynamics and tides can impact a planet's climate. In Section \ref{K186}, we study Kepler-186f, the first Earth-sized planet in the HZ of a cool star and discuss what can be said about its spin state. Finally, we conclude in Section \ref{conclusion}.   
  

\section{Tidal evolution}\label{tides}

The tidal model used here is a re-derivation of the equilibrium tide model of \citet{Hut:1981} as done in \citet{EKH1998}. We use the constant time lag model and consider non-coplanar situations as in \citet{Leconte2010}.

Let us consider a system with N planets orbiting a star. In order to compute the correct tidal evolution of the system, one has to consider the tide raised by the star on each planet (called the planetary tide) but also the tide raised by each planet on the star (called the stellar tide). 
The planetary tide and stellar tide act on different quantities. Depending on the body considered and on its tidal dissipation factor (a quantity linked to the constant time lag), the tides act on different timescales.

\subsection{Effect of the stellar tide}

The force created by the stellar tide will influence the semi-major axis, eccentricity and inclination of the orbit of the planet. The resulting torque will impact the rotation rate of the star. As the mass of the star is usually much higher than that of the planet, the planet cannot distort the star significantly and the corresponding evolutions will occur on long timescales.  

The semi-major axis evolution of the planet depends on the position of the planet with respect to the corotation radius, which is the orbital radius where the orbital period matches the star's rotation period. If the planet is inside the corotation radius, the planet migrates inwards and eventually falls on the star. If the planet is outside the corotation radius, it migrates outwards. The planet will affect the rotation rate of the star, however this effect is negligible in most cases. However, the rotation of a star can be significantly increased when a massive planet falls on it \citep{Bolmont2012}. 

Unless the star is a very fast rotator, the stellar tide acts to decrease the eccentricity. When the star is a very fast rotator, a slingshot effect can lead to an increase of the eccentricity. The stellar tide acts to decrease the inclination of the planet: it brings back the orbital plane into the equatorial plane of the star. 

 \subsection{Effect of the planetary tide}

The force created by the planetary tide will influence the semi-major axis and eccentricity of the orbit of the planet. The resulting torque will impact the rotation rate of the planet: the rotation period and the obliquity. The star can distort the planet significantly and in particular the evolution of the spin will occur on short timescales.  

Very quickly in the evolution of a circular orbit planet, the rotation will tend towards synchronization, which is a state for which the rotation period of the planet is equal to its orbital period. If the orbit is eccentric, the planet quickly reaches pseudo-synchronization, which means that its rotation tends to be synchronized with the orbital angular velocity at periastron \citep{Hut:1981}. The resulting rotation period is shorter than the orbital period. In the meantime, the obliquity of the planet quickly tends to zero.  

On longer timescales, after the pseudo-synchronization state is reached, the planetary tide makes the eccentricity and semi-major axis of the planet decrease. 
 
\section{Planets around brown dwarfs}\label{BDs}

Planets around brown dwarfs are interesting to study for three reasons: the HZ is located close-in \citep{Selsis2007,Andreeshchev2004}, so planets in the HZ experience strong tidal interactions and these HZ planets are easy to detect \citep{Belu2013,Triaud2013}. 

\subsection{One planet system}

\begin{figure}[ht!]
	\centering
	\includegraphics[width=0.8\textwidth,clip]{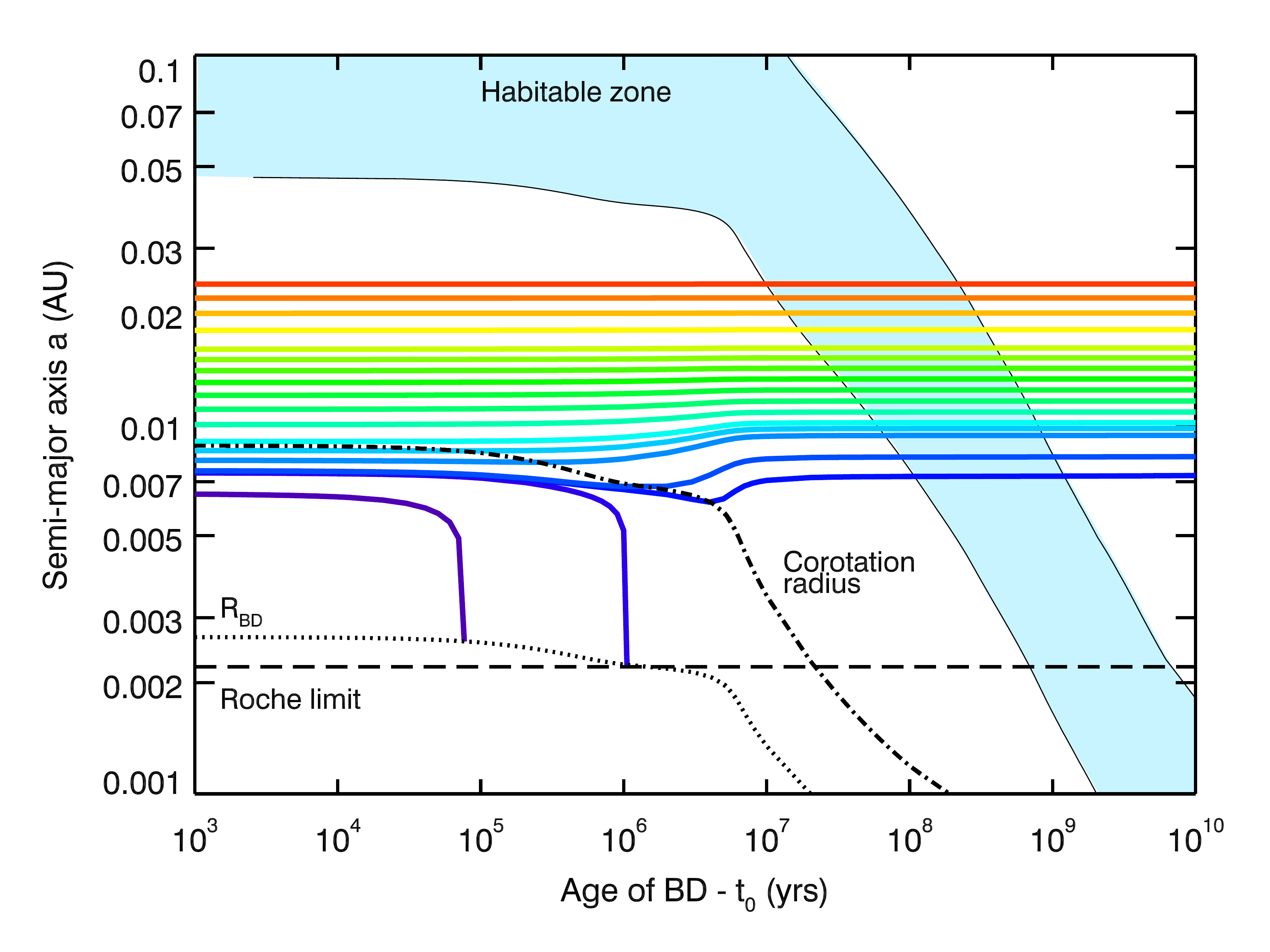}      
	\caption{Evolution of the orbital distance of an Earth-sized planet orbiting a BD of $0.04~M_\star$. The colored lines correspond to the evolution for different initial orbital distances. The black dashed dotted line corresponds to the corotation radius, the dotted line corresponds to the BD's radius. The insolation HZ is also plotted (blue region). The thick dashed line corresponds to the Roche limit. $t_0$ is the initial time, named ``time zero''. It corresponds to the time of protoplanetary disk dispersal, taken here as 1~Myr \citep{Bolmont2011}.}
	\label{fig1}
	\end{figure}

\citet{Bolmont2011} investigated the influence of tides on the orbital evolution of single-planet systems orbiting brown dwarfs (BDs) of different masses. The BD tide, the tide created by the planet in the BD, makes the planets either fall on the BD or migrate slightly outwards (see Figure \ref{fig1}). Not taking this migration into account would over-evaluate the time a planet spends in the HZ \citep[as was done in][]{Andreeshchev2004}. However \citet{Bolmont2011} showed that, despite this outward migration, planets around BDs more massive than $0.04~M_\star$ could stay in the HZ up to a few gigayears. 
	
Let us consider a single planet of $1~M_\oplus$ orbiting a $0.04~M_\star$ BD at 0.01~AU. It would reach the HZ about $100$~Myr after the protoplanetary disk dispersal \citep{Bolmont2011}. A hundred million years is enough to erode the obliquity of the planet and bring it in pseudo-synchronization. If the eccentricity is very small the pseudo-synchronization rate is very similar to the synchronization rate (a difference of $3.5$\% for an eccentricity of $0.06$ and of $0.1$\% for an eccentricity of $0.01$). 
Therefore, when the planet reaches the HZ, it is likely to have a null obliquity and a synchronous rotation. 

This means that this planet always shows the same side to the BD and that its poles receive very little light. This raises the problem of the possible existence of cold traps: i.e. regions on the planet where the temperature is constantly lower than $273$~K and where all the water of the planet will condense \citep{Joshi1997,Joshi2003}. 
In this configuration, night side and poles could be cold traps and although the planet is in the HZ, it would not be able to host surface liquid water. 
However, this situation is not completely hopeless, a dense enough atmosphere that would allow the repartition of heat around the whole planet could prevent cold traps from capturing the water content of the planet. \citet{Wordsworth2011} showed that Gliese~581~d \citep[if it exists;][]{Robertson2014} could have a habitable climate despite a synchronous rotation state. 

\subsection{Multiple planet system}

The situation differs if the planet is part of a multiple planet system. Because of planet-planet interactions, eccentricity and obliquity do not tend to 0 but an equilibrium value which is the result of the competition between tidal damping and planet-planet excitation. 

An extreme case of planet-planet excitation is the mean motion resonance (MMR). When two planets are in MMR, the ratio between their orbital periods is commensurable and the eccentricity is excited to higher levels. A close-by example is the 1:2:4 MMR between the 3 inner satellites of Jupiter (Io, Europa, Ganymede). This eccentricity excitation maintained by the resonance causes Io to experience an intense internal heating due to the stress it experiences on one orbit. This internal heating is estimated at about $3$~W/m$^2$ \citep[e.g.,][]{Spencer2000}, which is about 40 times higher than the internal heat flux of Earth \citep[about $0.08$~W/m$^2$ and mainly due to radioactivity; e.g.,][]{Davies2010}. On Io this heat flux gives rise to intense volcanic activity. A tidally evolving planet in the HZ could be heated up and driven in a runaway greenhouse state \citep[thus creating a ``tidal Venus'';][]{Barnes2013}.

\medskip

Let us consider the case of three Earth-sized planets orbiting just outside the corotation radius of a $0.08~M_\star$ BD. We consider a) a BD with a tidal dissipation factor similar to hot Jupiters' \citep[due to their likeliness in structure, we consider here that BDs and hot Jupiters have a similar dissipation factor][]{Hansen2010} and b) a BD with a tidal dissipation of ten times the dissipation of hot Jupiters. 

\begin{figure}[ht!]
	\centering
	\includegraphics[width=0.9\textwidth,clip]{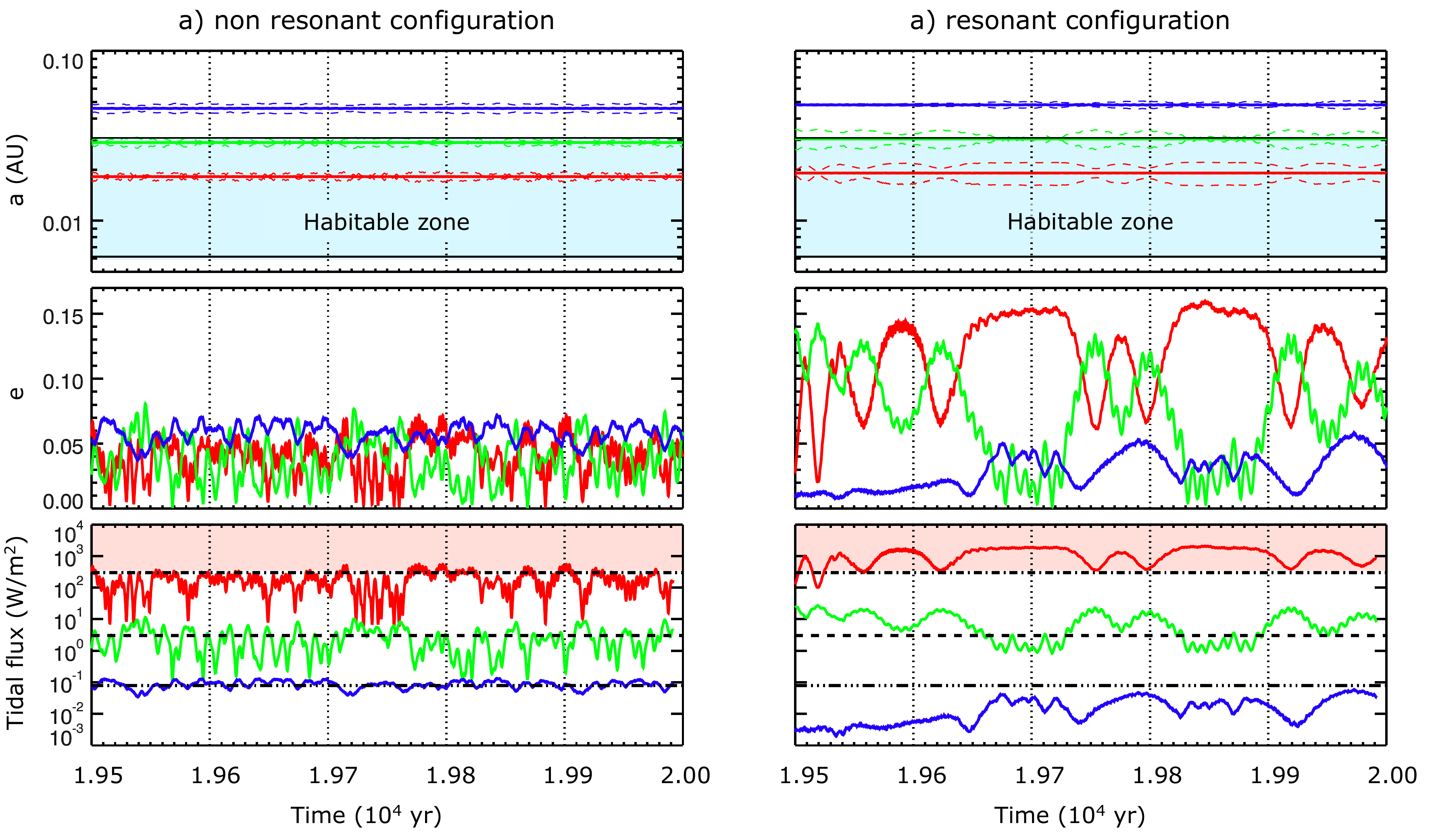}      
	\caption{Evolution of the orbital distance, eccentricity and tidal heat flux of three Earth-sized planets orbiting a BD of $0.08~M_\star$ in a) a non-resonant configuration, b) a resonant configuration (1:2 MMR). Top graph: the full colored lines correspond to the semi-major axis evolution of the 3 planets, and the dashed lines correspond to their perihelion and aphelion distances. The blue shaded region is the HZ. Middle graph: eccentricity of the 3 planets. Bottom graph: the full colored lines correspond to the tidal heat flux of the 3 planets. The black dashed dotted line corresponds to $300$~W/m$^2$, the dashed line corresponds to Io's heat flux and the dashed 3 dots line corresponds to Earth's heat flux. The shaded red region corresponds to where the heat flux is so high that the planet is in a runaway greenhouse state.}
	\label{fig2}
	\end{figure}

In both cases, the planets experience a convergent outward migration but they will do so quicker in case b) than a). The difference in dissipation will lead to two dynamically different systems. As the migration is faster in case b) planets enter a MMR chain (1:2:4) after a few million years of evolution. However in case a), they stay outside of resonance. 
Figure \ref{fig2} shows the evolution of such a system at an older age (1~Gyr), the HZ has shrunk and the two inner planets are now in the HZ. The eccentricities of the planets in case a) are relatively small $<0.07$ but in case b) due to the MMR excitation, they can reach $0.15$. 
 
Figure \ref{fig2} also shows the evolution of the tidal heat flux for each planet, compared to three values: the limit of runaway greenhouse $300$~W/m$^2$ \citep{Kopparapu2013}\footnote{Note that this limit has been re-evaluated with 3D climate simulations at 375~W/m$^2$ by \citet{Leconte2013Nat}}, Io's tidal heat flux and Earth's heat flux. 

In case a), the average tidal heat flux of the inner planet (red) is below the limit of $300$~W/m$^2$ and sometimes exceeds this limit for a few $10$~yr. If we consider that the tidal heat flux is the only heat source of the atmosphere of the planet (we neglect momentarily the insolation from the BD), we could say that most of the time the inner planet can have habitable conditions. 
Let us consider that this planet has oceans. When the tidal heat flux exceeds the runaway greenhouse limit, the oceans will start evaporating. However, for an Earth-like ocean, 10 years would not be enough to evaporate it all, so that the water reservoir would survive. 

However, in case b), due to the excitation of the eccentricity of the orbits, the tidal heat flux of the inner planet always exceeds the runaway greenhouse limit. Thus, this HZ planet would be too hot to host surface liquid water. 

The middle planet (in green) is on the outer edge of the HZ, and its tidal flux is relatively low (in average it is about Io's tidal flux). For case a), this planet could potentially have habitable conditions\footnote{We don't discuss here the influence of an intense volcanic activity on habitability \citep{Hanslmeier2012}.}. For case b), the planet spends some time around aphelion outside the insolation HZ and it has a tidal heat flux a bit superior to Io's in average. Without tidal heating, this planet could be too cold to be able to sustain a potential liquid water reservoir, however taking into account tidal heating will improve the conditions for habitability. One could imagine a more extreme case of a planet on an orbit completely outside the HZ but heated up by tides sufficiently to be able to host surface liquid water.

When assessing habitability of planets in the HZ of BDs, one should investigate if tides are strong enough to drive a runaway greenhouse. If the atmosphere receives an average flux ($\Phi_\star + \Phi_{{\rm tides}}$) lower than $300$~W/m$^2$, the planet can sustain a liquid water reservoir, but if it receives an average flux higher than $300$~W/m$^2$ the planet will be too hot to be able to sustain a liquid water reservoir.


\section{Kepler-186}\label{K186}

The planetary tide influences the rotation of the planet, which is an important parameter for climate studies. We take here the example of the planet Kepler-186f, the first Earth-sized planet in the HZ of a star \citep{Quintana2014}. Using the system parameters given in \citet{Bolmont2014}, we compute the tidal evolution of the system and focus our attention on the fifth planet. 

\begin{SCfigure}
	\centering
	\caption{Long-term evolution of the obliquity (top) and rotation period (bottom) of Kepler-186f (set $\mathcal{A}$). Each set of linestyle curves represents a different initial spin rate and each set of colored curves represents a different initial obliquity. The thick, black dashed line represents the pseudo-synchronous rotation which, for this zero-eccentricity example, is the 1:1 spin--orbit resonance. This Figure comes from \citet{Bolmont2014}.}
	\includegraphics[width=0.4\textwidth]{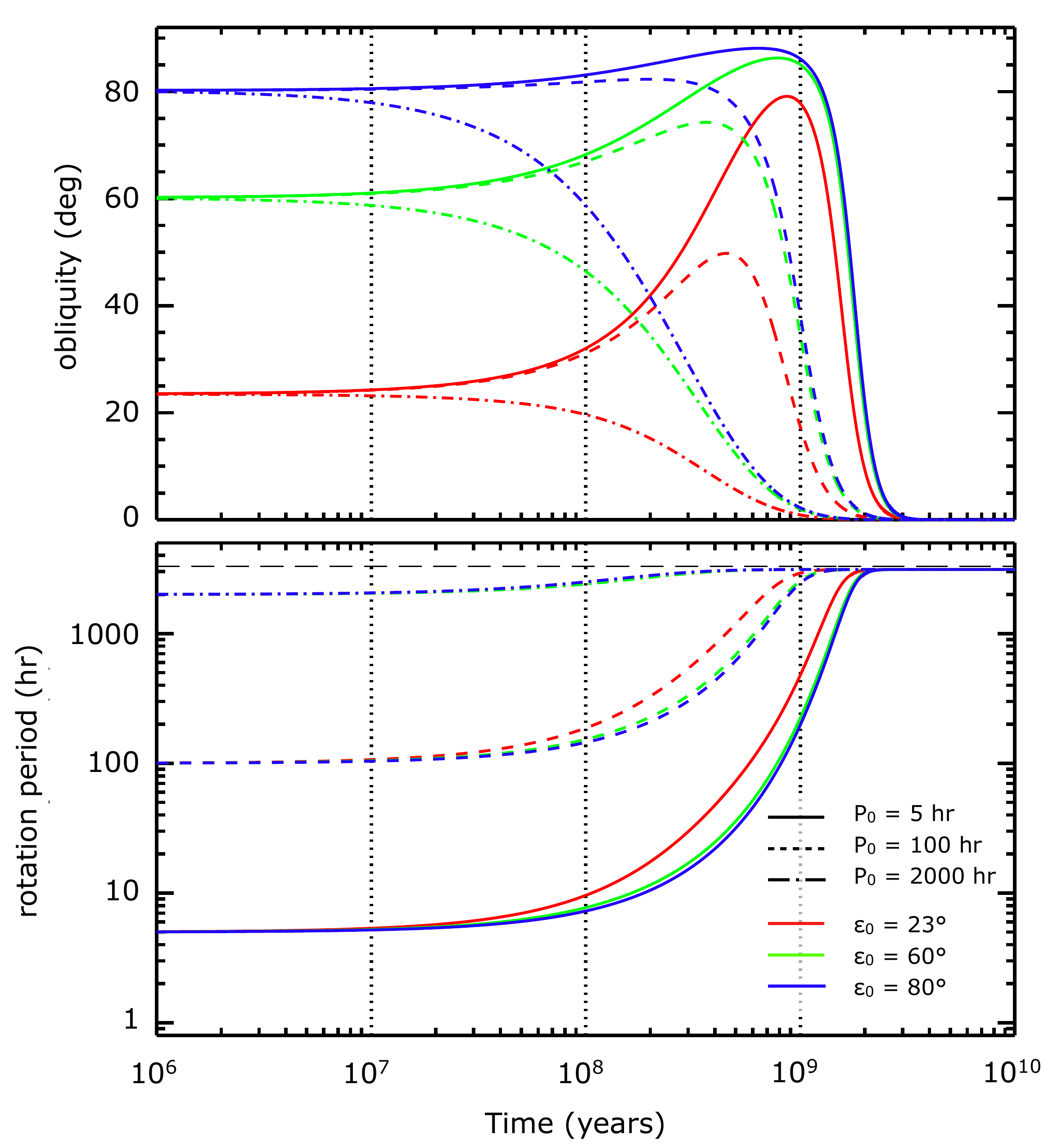}
	\label{fig3}
	\end{SCfigure}

Contrary to the four inner planets, which quickly reach a pseudo-synchronous state and low obliquities (in $\lesssim 1$~Myr), the evolution timescale of the rotation of Kepler-186f is much longer. Figure \ref{fig3} shows the evolution of obliquity and rotation period for Kepler-186f for different initial obliquities and rotation periods. The initial values are of course not known but N-body simulations of terrestrial accretion tend to produce planets with fast initial spins and isotropically distributed obliquities \citep{kokubo07}.  

Figure \ref{fig3} shows that Kepler-186f's obliquity increases for all but the slowest initial spin rate. The obliquity increases for a few hundred megayears and this period is followed by a long, slow decay that lasts 2-3 Gyr (for the arbitrarily chosen range of initial spin rates and an Earth-like tidal dissipation). The equilibrium obliquity of the planet is very low, it does not feel the dynamical interactions of the four inner planet.

Given the estimated age of the system \citep[$\gtrsim$4~Gyr according to][]{Quintana2014}, Kepler-186f could be in pseudo-synchronous rotation with a very small obliquity (see Fig. \ref{fig3}). But if the system is a bit younger (1~Gyr), or if the planet dissipates less than what was assumed, Kepler-186f could have a faster rotation and a very high obliquity ($\lesssim 80$ degrees).

This would impact the climate of the planet: a non-negligible obliquity causes seasonal effects while a negligible obliquity causes a low insolation at the poles; the rotation has an impact on the heat transport in the atmosphere and if it is sufficiently close to the synchronous rotation, the night side could a cold trap. It is therefore necessary to consider all possible rotation states when assessing the climate and habitability of this planet. 


\section{Conclusions}\label{conclusion}

The climate of a planet depends on many parameters. It depends on the atmospheric pressure, temperature and composition but also on astronomic quantities: orbital distance (insolation), eccentricity of the orbit (insolation, seasons), rotation of the planet (cold trap, heat redistribution), obliquity of the planet (cold trap, seasons), tidal heat flux (acting as an increase in insolation). The stellar tide influences the orbital distance and the eccentricity of the planet while the planetary tide influences all these quantities. There is therefore a very tight link between the dynamics of tidal evolution and the climate of a planet. 

\begin{acknowledgements}
S. N. R. and F. S. acknowledge support from the Programme National de Plan\'etologie (PNP). 

\end{acknowledgements}



%
\end{document}